\documentclass[twocolumn,showpacs,superscriptaddress,nofootinbib,floatfix,preprintnumbers]{revtex4-1}

\usepackage[english]{babel}
\usepackage[utf8]{inputenc}
\usepackage{amsfonts,amsmath,amssymb,amsbsy,amsthm,bbold,bm,mathtools,fixmath,slashed}
\usepackage{comment,enumerate,footnote,graphicx,subfloat,relsize,footnote,eurosym}
\usepackage{array,tabularx,tabu,multirow,empheq}
\usepackage[usenames,dvipsnames]{xcolor}

\newcommand{\Tr}{\text{Tr}}  

\usepackage{cleveref}
\crefformat{pluralequation}{#2\black{eqs.~(}#1\black{)}#3}
\Crefformat{pluralequation}{#2\black{Equations~(}#1\black{)}#3}
\crefformat{pluralfigure}{#2\black{figs.~}#1#3}
\Crefformat{pluralfigure}{#2\black{Figures~}#1#3}

\usepackage{tikz}
\usetikzlibrary{arrows,shapes}
\usetikzlibrary{trees,patterns}
\usetikzlibrary{matrix,arrows} 				
\usetikzlibrary{positioning}				  
\usetikzlibrary{calc,through}				  
\usetikzlibrary{decorations.pathreplacing}  
\usepackage{pgffor}							

\usetikzlibrary{decorations.pathmorphing}	
\usetikzlibrary{decorations.markings}
\tikzset{
>=stealth', 
vector/.style={decorate, decoration={snake}, draw},
provector/.style={decorate, decoration={snake,amplitude=2.5pt}, draw},
antivector/.style={decorate, decoration={snake,amplitude=-2.5pt}, draw},
bigvector/.style={decorate, decoration={snake,amplitude=4pt}, draw},
fermion/.style={draw=black, postaction={decorate},
decoration={markings,mark=at position .55 with {\arrow[draw=black]{>}}}},
fermionbar/.style={draw=black, postaction={decorate},
decoration={markings,mark=at position .55 with {\arrow[draw=black]{<}}}},
fermionnoarrow/.style={draw=black},
gluon/.style={decorate, draw=black,
decoration={coil,amplitude=4pt, segment length=5pt}},
scalar/.style={dashed,draw=black, postaction={decorate},
decoration={markings,mark=at position .55 with {\arrow[draw=black]{>}}}},
scalarbar/.style={dashed,draw=black, postaction={decorate},
decoration={markings,mark=at position .55 with {\arrow[draw=black]{<}}}},
scalarnoarrow/.style={dashed,draw=black},
momentum/.style={draw=black, postaction={decorate},
decoration={markings,mark=at position 1 with {\arrow[draw=black]{>}}}},
antimomentum/.style={draw=black, postaction={decorate},
decoration={markings,mark=at position 0.1 with {\arrow[draw=black]{<}}}}
}

\tikzstyle{block} = [draw, rectangle, 
    minimum height=3em, minimum width=6em]
    
\newcommand{\nc}{\newcommand}
\nc{\pd}{\partial}
\nc{\bea}{\begin{eqnarray}}
\nc{\eea}{\end{eqnarray}}
\nc{\bal}{\begin{alignedat}}
\nc{\eal}{\end{alignedat}}
\nc{\beq}{\begin{equation}}
\nc{\eeq}{\end{equation}}
\nc{\bit}{\begin{itemize}}
\nc{\eit}{\end{itemize}}
\nc{\benu}{\begin{enumerate}}
\nc{\eenu}{\end{enumerate}}
\nc{\bdes}{\begin{description}}
\nc{\edes}{\end{description}}
\nc{\bma}{\begin{pmatrix}}
\nc{\ema}{\end{pmatrix}}

\newcommand{\black}[1]	{{\color{black} #1}}
\colorlet{lightblue}{blue!30}
\colorlet{lightred}{red!30}

\nc{\nn}{\nonumber}
\nc{\hc}{\text{h.c.}}
\nc{\cc}{\text{c.c.}}

\nc{\abs}[1]{\left| #1 \right|}
\def\[{\left[}
\def\]{\right]}
\def\({\left(}
\def\){\right)}
\def\<{\langle}
\def\>{\rangle}

\def\g5{\gamma_{5}}

\def\g{\gamma}

\def\aD			{\alpha}

\def\BSF		{\mathsmaller{\rm BSF}}
\def\bBSF		{\mathsmaller{\rm bBSF}}
\def\mBSF		{\mathsmaller{\rm mBSF}}

\def\mV			{m_{\mathsmaller{V}}}

\def\vrel		{v_{\rm rel}}

\newcommand{\bra}[1]{ \langle {#1} | }
\newcommand{\ket}[1]{ | {#1} \rangle }

\begin{document}

\title{Rapid bound-state formation of Dark Matter in the Early Universe}

\author{Tobias Binder}
\email{tobias.binder@ipmu.jp}
\affiliation{Kavli IPMU (WPI), UTIAS, The University of Tokyo, Kashiwa, Chiba 277-8583, Japan}

\author{Kyohei Mukaida}
\email{kyohei.mukaida@desy.de}
\affiliation{Deutsches Elektronen-Synchrotron (DESY), Notkestra{\ss}e 85, Hamburg, D-22607 Germany}

\author{Kalliopi Petraki}
\email{kpetraki@lpthe.jussieu.fr}
\affiliation{Sorbonne  Universit\'e,  CNRS,  Laboratoire  de  Physique  Th\'eorique  et  Hautes  Energies, LPTHE, F-75252 Paris, France}
\affiliation{Nikhef, Science Park 105, 1098 XG Amsterdam, The Netherlands}

\date{\today}

\begin{abstract} 
The thermal decoupling description of dark matter (DM) and co-annihilating partners is reconsidered. If DM is realized at around the TeV-mass region or above, even the heaviest electroweak force carriers could act as long-range forces, leading to the existence of meta-stable DM bound states. The formation and subsequent decay of the latter further deplete the relic density during the freeze-out process on top of the Sommerfeld enhancement, allowing for larger DM masses. While so far the bound-state formation was described via the emission of an on-shell mediator ($W^{\pm}$, $Z$, $H$, $g$, photon or exotic), we point out that this particular process does not have to be the dominant scattering-bound state conversion channel in general. If the mediator is coupled in a direct way to any relativistic species present in the Early Universe, the bound-state formation can efficiently occur through particle scattering, where a mediator is exchanged virtually. To demonstrate that such a virtually stimulated conversion process can dominate the on-shell emission even for all temperatures, we analyze a simplified model where DM is coupled to only one relativistic species in the primordial plasma through an electroweak-scale mediator. We find that the bound-state formation cross section via particle scattering can exceed the on-shell emission by up to several orders of magnitude.
\end{abstract}

\maketitle
\preprint{IPMU19-0148} 
\preprint{DESY 19-181} 
\preprint{Nikhef-2019-049}


\section{Introduction \label{Sec:intro}}

One of the leading dark matter hypothesis is that it consists of Weakly-Interacting-Massive-Particles (WIMPs)~\cite{Lee:1977ua, Ellis:1983ew, Arcadi:2017kky}, which can explain the observed DM abundance~\cite{Ade:2015xua} in a natural way through the thermal production mechanism. While strong upper bounds on the coupling strength of the WIMP to Standard Model (SM) particles, derived from direct, indirect and collider searches, rule out many electroweak mass-scale realizations in the thermal production scenario, the TeV mass region and above still remains an attractive and much less constrained possibility. 

The heavier WIMPs are, the more important it is to include quantum mechanical effects induced by long-range interactions in the thermal decoupling description for predicting the relic abundance accurately. In a seminal work~\cite{Hisano:2006nn}, it has been pointed out that already around a three TeV mass, it is possible that even the heaviest massive gauge bosons of the SM effectively act as attractive long-range forces between a WIMP pair, leading to an enhanced annihilation cross section during the freeze-out process. This so-called \emph{Sommerfeld enhancement} (SE) \cite{doi:10.1002/andp.19314030302} or \emph{Sakharov enhancement} \cite{Sakharov:1948yq,Belotsky:2004st} lowers the predicted thermal relic abundance compared to a tree-level computation, which in turn allows for larger masses of the annihilating DM particles to compensate for the effect (see, e.g., ~\cite{Freitas:2007sa,Cirelli:2007xd,Berger:2008ti,Hryczuk:2010zi,Hryczuk:2011tq,Harigaya:2014dwa,ElHedri:2016onc,Beneke:2016jpw,Beneke:2016ync} and~\cite{Slatyer:2009vg,Beneke:2014gja,Blum:2016nrz,Braaten:2017gpq,Braaten:2017dwq,Braaten:2017kci,Tang:2018viw} for formal aspects).

Further quantum mechanical effects caused by attractive long-range interactions are bound-state solutions in the two-particle spectrum of WIMPs~\cite{Pospelov:2008jd,MarchRussell:2008tu,Shepherd:2009sa}. The formation of these bound states and their subsequent decay into SM particles additionally depletes the relic density, allowing for even heavier masses~\cite{vonHarling:2014kha}. For every SM force-carrier, possible DM scenarios have been found in the literature where the inclusion of the SE and bound-state formation ({\bf BSF}) is relevant. Famous examples in Supersymmetric extensions of the SM are cases where DM co-annihilates with closely mass degenerate partners, which are either electroweakly or color charged~\cite{Ellis:2015vna,Liew:2016hqo,Asadi:2016ybp,Johnson:2016sjs,Kim:2016zyy,Mitridate:2017izz,Harz:2018csl,Biondini:2018pwp,Biondini:2018ovz,Fukuda:2018ufg,Biondini:2019int}. In addition to the photon, gluon, $Z$ or $W^{\pm}$ boson induced bound states in mentioned cases, also the Higgs particle~\cite{Harz:2017dlj,Biondini:2018xor} can attractively contribute to confine DM into a meta-stable bound state, even for otherwise repulsive color octet states~\cite{Harz:2019rro}. Furthermore, these quantum mechanical phenomena certainly play a role in bottom-up motivated~\cite{Kamada:2016euw,Kaplinghat:2019dhn} scenarios, e.g., Self-Interacting DM~\cite{Tulin:2017ara} with light mediators~\cite{Buckley:2009in,Aarssen:2012fx,Tulin:2013teo,Bringmann:2016din,An:2016gad,Baldes:2017gzw,Baldes:2017gzu,Kamada:2018zxi,Kamada:2018kmi,Matsumoto:2018acr,Kamada:2019jch,Ko:2019wxq}, where long-range interactions are introduced by assumption. 

So far, the DM bound-state formation process in the Early Universe was described via the emission of an on-shell mediator. While this resembles the situation of SM neutral hydrogen recombination in the matter-dominated epoch, an interesting question is by how much scattering events with primordial plasma constituents in the radiation dominated epoch could additionally stimulate the conversion process through virtual mediator exchanges.

In the case of annihilating heavy quarkonia, produced in a quark-gluon plasma (QGP) in heavy ion collision at, e.g., the Large-Hadron-Collider, it is well known that the dissociation of the heavy quarkonia bound states via the absorption of an on-shell gluon is not the dominant process for temperature larger than the binding energy (see, e.g.,~\cite{Brambilla:2013dpa,Hong:2018vgp}). Instead it is the dissociation via light quark and gluon scattering which dominates in that temperature regime. By the argument of detailed balance, this automatically implies that the dominant heavy quarkonia BSF channel must be via light parton scattering as well. Since the system of heavy quarks inside a QGP is similar to DM in the Early Universe (where the latter contains many more relativistic species), this insight might have already profound implications for DM models where light or massless mediators in co-annihilation scenarios are involved.

In this work, we investigate a mediator whose mass is of the order of the $Z,W^{\pm},$ or Higgs boson mass. The latter cases might be of particular interest, since the kinetic energy of WIMPs during the freeze-out process could be not enough to emit an on-shell massive boson at the electroweak scale in order to form a bound state. For such a case, it is known that BSF has only marginal effects due to the kinematical block. However, inside a relativistic plasma background, bath-particle scattering can stimulate the BSF process by inducing a mediator excitation virtually. The virtual exchange implies that BSF via bath-particle scattering has no kinematical block and might entirely dominate over the on-shell mediator emission.

As a proof of concept, this article investigates for the first time this possibility in a simplified DM model, which is aimed to parametrically resemble more realistic cases with $Z,W^{\pm},H$ or other exotic interactions. The details of the model and the computation based on a generalized BSF cross section~\cite{Binder:DivCancel}, capturing higher-order processes in a proper thermal field theory framework, are shared in section~\ref{sec:crosssections}. The numerical evaluation of the thermally averaged quantities and the implications of strongly enhanced bound-state formation rates are discussed in section~\ref{sec:impl}. Section~\ref{sec:con} concludes this work.


\section{Model and cross sections}
\label{sec:crosssections}

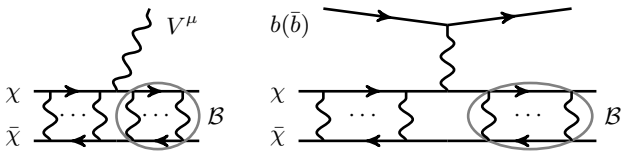
\begin{figure}[t!]
\centering
\begin{tikzpicture}[line width=1pt, scale=1.1]
\begin{scope}[shift={(0,0)}]
\node at ( 0.8, 0.8){$V^\mu$};
\node at (-1.25,-0.05)  {$\chi$};
\node at (-1.25,-0.55){$\bar{\chi}$};
\draw[fermion]	(-1, 0)   -- (0, 0);	\draw[fermion]	(0, 0)   -- (1, 0);
\draw[fermion]	(0,-0.6) -- (-1,-0.6);	\draw[fermion]	(1,-0.6) -- (0,-0.6);
\draw[fill=none,gray] (0.5,-0.3) ellipse (0.5 and 0.4);
\node at (1.2,-0.3){${\cal B}$};
\draw[vector]	(-0.8,-0.6) -- (-0.8,0);
\node at (-0.5,-0.3){$\cdots$};
\draw[vector]	(-0.2,-0.6) -- (-0.2,0);
\draw[vector]	( 0.2,-0.6) -- ( 0.2,0);
\node at (0.5,-0.3){$\cdots$};
\draw[vector]	( 0.8,-0.6) -- (0.8,0);
\draw[vector]	(0,0) -- (0.4,1);
\end{scope}
\begin{scope}[shift={(4,0)}]
\node at (-1.9, 0.8){$b(\bar{b})$};
\node at (-2.05,-0.05) {$\chi$};
\node at (-2.05,-0.55) {$\bar{\chi}$};
\draw[fermion]	(-1.5, 1) -- (0, 0.8);		\draw[fermion]	(0, 0.8) -- (1.5, 1);
\draw[fermion]	(-1.8, 0) -- (0, 0);		\draw[fermion]	(0, 0) -- (1.8, 0);
\draw[fermion]	(0,-0.6) -- (-1.8,-0.6);	\draw[fermion]	(1.8,-0.6) -- (0,-0.6);
\draw[fill=none,gray] (1,-0.3) ellipse (0.75 and 0.4);
\node at (2,-0.3){${\cal B}$};
\draw[vector]	(-1.5,-0.6) -- (-1.5,0);
\node at (-1,-0.3){$\cdots$};
\draw[vector]	(-0.5,-0.6) -- (-0.5,0);
\draw[vector]	( 0.5,-0.6) -- ( 0.5,0);
\node at (1,-0.3){$\cdots$};
\draw[vector]	( 1.5,-0.6) -- (1.5,0);
\draw[vector]	(0,0) -- (0,0.8);
\end{scope}
\end{tikzpicture}
\caption{\label{fig:FeynmanDiagrams} 
Feynman diagrams for bound-state formation via on-shell mediator emission (left) and via bath-particle scattering (right) shown. Both processes contain also the diagram where the mediator is attached to $\bar{\chi}$.
}
\end{figure}

We consider a rather simple three-parameter model:
\begin{align}
{\cal L} \supset -g \bar{\chi}\gamma^{\mu}\chi V_{\mu} - g \bar{b}\gamma^{\mu}b V_{\mu},
\label{eq:L}
\end{align}
where non-relativistic DM $(\chi)$ with mass $m_{\chi}$ and the ultra-relativistic primordial plasma particles ($b$) are Dirac Fermions. The mass of the Abelian vector mediator is assumed to fulfill $m_V \lesssim \alpha m_{\chi}$, where $\aD \equiv g^2 / (4\pi)$ is the fine structure constant, such that bound-state solutions exist in the DM two-particle spectrum.

The generalized bound-state formation cross section, which includes the on-shell emission and the bath-particle scattering in fig.~\ref{fig:FeynmanDiagrams}, as well as other higher order processes in a proper thermal field theoretical framework, is given by~\cite{Binder:DivCancel} (see also App.~\ref{app:bsf}):
\begin{align}
\sigma_{nlm}^{\BSF} v_{\text{rel}} = \! \! \int \! \!  \frac{\text{d}^3 p}{(2\pi)^3} D^{-+}_{\mu \nu}(P) \! \!\sum_{\text{spins}} \! {\cal T}^{\mu}_{\mathbf{k}, nlm} (P) {\cal T}^{\nu \star}_{\mathbf{k},nlm} (P). \label{eq:generalxsection}
\end{align} 
$P=(P^0=\Delta E,\mathbf{p})$ is the four momentum of the vector boson $V$, whose energy component is fixed by the positive quantity $\Delta E \equiv {\cal E}_{\bf k} - {\cal E}_{nlm}$. $\Delta E$ is the total energy emitted in the inelastic conversion, i.e., relative kinetic energy ${\cal E}_{\bf k} = {\bf k^2}/(2\mu) = \mu \vrel^2/2$ of DM with reduced mass $\mu$, plus the absolute value of the negative binding energy ${\cal E}_{nlm}$ of the bound state with quantum numbers $nlm$. The central Eq.~(\ref{eq:generalxsection}) consist of a two-point correlation function $D^{-+}_{\mu \nu}$ of the vector boson, which is contracted with non-relativistic (NR) scattering-bound state transition matrix elements at the Born-level.
The integral over $\mathbf{p}$ picks out the various physical poles appearing in $D^{-+}_{\mu \nu}$.

The matrix elements are defined in dipole approximation (dip) as:
\begin{align}
{\cal T}^{\mu}_{\mathbf{k},nlm} (P) \equiv (g_{\chi} g_{\bar{\chi}} 4 m_{\chi}^2 2M)^{-1/2} \mathcal{M}^{\mu}_{\mathbf{k},nlm} \big|_{\text{dip}}^{\text{NR}}\;, \label{eq:tdef}
\end{align}
where $M$ is the bound-state mass and initial DM spin d.o.f. are $g_{\chi} g_{\bar{\chi}}=4$. $\mathcal{M}^{\mu}$ is defined as the Fourier transform of the $\mathcal{S}$cattering-$\mathcal{B}$ound state transition matrix element of the current operator, here given by:
\begin{align}
\!\!\slashed{\delta}^4\mathcal{M}^{\mu}_{\mathbf{k},nlm}= \!\!\int \! \text{d}^4 x\; e^{iPx} \bra{\mathcal{B}_{nlm}} g \bar{\chi}(x)\gamma^{\mu} \chi(x) \ket{\mathcal{S}_{\mathbf{k}}}. \label{eq:vertex}
\end{align}
$\slashed{\delta}^4 =(2\pi)^4 \delta^4$ is the four-momentum conserving delta function. Spin indices are implicit, see  App.~\ref{app:states} for our convention of the states.

For the rest of this work, the dominant direct capture into the ground state is considered. Contracting the spinors in Eq.~(\ref{eq:vertex}), taking the non-relativistic and dipole limit, we obtain~\cite{Petraki:2016cnz}:
\begin{align}
\! \! \! \sum_{\text{spins}} {\cal T}^{\mu}_{\mathbf{k},100} {\cal T}^{\nu \star}_{\mathbf{k},100} =  
 \frac{4 \pi h(\zeta,\xi)}{\mu^{3} \Delta E^2} \begin{pmatrix} \mathbf{p} \cdot  \hat{\mathbf{k}} \\  \Delta E \; \hat{\mathbf{k}}   \end{pmatrix}^\mu \begin{pmatrix} \mathbf{p} \cdot  \hat{\mathbf{k}} \\  \Delta E \; \hat{\mathbf{k}}   \end{pmatrix}^\nu ,\label{eq:Tgroundstate}
\end{align}
with $\Delta E = {\cal E}_{\bf k} + |{\cal E}_{100}|$ for the ground state, unit relative-momentum vector $\hat{\mathbf{k}}$, and we defined $\zeta\equiv\alpha/v_{\text{rel}}$ and $\xi\equiv\alpha m_{\chi}/(2m_V)$. Note that once the Born transition matrix element Eq.~(\ref{eq:vertex}) is determined [e.g., Eq.~(\ref{eq:Tgroundstate})] it can be reused in Eq.~(\ref{eq:generalxsection}) at any order of the perturbative expansion of the two-point correlation function. As a check, $\cal T$ fulfills the current conservation $P_{\mu}T^{\mu}_{\mathbf{k},100}=0 $ as a consequence of the global symmetry. Dimensionless $h(\zeta,\xi)$ contains the dipole approximation of the overlap integral, which can be computed only numerically for a massive mediator~\cite{Petraki:2016cnz}, see App.~\ref{app:states} for convention.

The two-point correlation function in coordinate space $D^{-+}_{\mu \nu}(x,y)\equiv \langle V_{\mu}(x)V_{\nu}(y)\rangle$, where $ \langle ... \rangle= \Tr[e^{- H_{\text{env}}/T} ...]$, encodes \emph{all} interactions with the primordial plasma environment and hence the information of the bath-particles. It is related to the \emph{spectral function} $D^{\rho}$ by the Kubo-Martin-Schwinger relation (see, e.g.,~\cite{Binder:2018znk}):
\begin{align}
D^{-+}_{\mu \nu}(\Delta E,\mathbf{p})= \left[1+f^{\text{eq}}_V(\Delta E)\right] D^{\rho}_{\mu \nu}(\Delta E,\mathbf{p}).\label{eq:KMS}
\end{align}
The equilibrium phase-space distribution $f^{\text{eq}}_V$ obeys Bose-Einstein statistics.
It is convenient to compute the spectral function from the retarded correlator via $D_{\mu \nu}^{\rho}= 2 \Im \left[i D^R_{\mu \nu} \right]$, since the latter obeys also in thermal field theory the Dyson-Schwinger equation, given in momentum space by
\begin{align}
D^R_{\mu \nu} = D^{R,0}_{\mu \nu} + D^{R,0}_{\mu \alpha} \Pi^{\alpha\beta}_{R} D^{R,0}_{\beta \nu} + ... \;.\label{eq:Dyson}
\end{align}
In the following, we show that BSF via on-shell mediator emission is reproduced from the free term, while BSF via bath-particle scattering is contained in the interaction term with the retarded self-energy $\Pi_R$. For both processes, Eq.~(\ref{eq:Tgroundstate}) can be reused to calculate the BSF cross section Eq.~(\ref{eq:generalxsection}), thanks to the factorization.

Although a self-energy with fermions is analyzed in this work, the formalism can capture non-abelian interactions, e.g. gluon scattering (triple vertex). For non-Abelian theories, ref.~\cite{Harz:2018csl} provides many expressions of the transition element Eq.~(\ref{eq:tdef}), ready to be used with Eq.~(\ref{eq:generalxsection}). For scalar mediators, one can drop ``$\mu \nu$" everywhere and adjust Eq.~(\ref{eq:vertex}) to Yukawa interactions.

\subsection{Capture via on-shell mediator emission}
To the lowest order in perturbation theory the retarded propagator of the massive vector mediator is given by $D^{R,0}_{\mu \nu}(P)=-ig_{\mu \nu}[P^2 -m_V^2 + i \text{sgn}(P^0)\epsilon ]^{-1}$. Taking $2\Im [i ...]$ of the latter to compute the spectral function, and inserting Eq.~(\ref{eq:KMS}) together with  Eq.~(\ref{eq:Tgroundstate}) into the cross section, the standard result for capture into the ground state via the on-shell emission of a massive vector mediator ({\bf{mBSF}}) is reproduced after performing elementary integrals (cf. Eq.3.7a in~\cite{Petraki:2016cnz}):
\begin{align}
\!\!\!\sigma^{\mBSF}_{100} v_{\text{rel}} \! =  \!  \frac{4  h(\zeta,\xi)\Delta E}{3 \mu^3}  \!\left[1+f^{\text{eq}}_V(\Delta E)\right] \! \frac{s_{\text{ps}}^{1/2}(3-s_{\text{ps}})}{2}, \label{eq:xsectionmbsf}
\end{align}
where $s_{\text{ps}}\equiv 1-m_V^2/\Delta E^2$ implies the kinematical suppression.
\subsection{Capture via bath-particle scattering}
The retarded self-energy for ultra-relativistic fermionic particles is given by (see App.~\ref{app:selgenergy}):
\begin{align}
\! \! \!  &\Pi^{R}_{\mu \nu}(\Delta E, \mathbf{p})=  g^2 \int \text{d}\Pi_1 \text{d}\Pi_2  \Tr[\gamma_{\mu} \slashed{P}_1 \gamma_{\nu} \slashed{P}_2] \times \nonumber \\
\! \! \! \bigg\{ \! &  \frac{i 2 \slashed{\delta}^3(\mathbf{p}+\mathbf{p}_1-\mathbf{p}_2)}{\Delta E +|\mathbf{p}_1|-|\mathbf{p}_2| + i \epsilon} \! \left[ f^{\text{eq}}_b(|\mathbf{p}_1|) \! - \! f^{\text{eq}}_{b}(|\mathbf{p}_2|) \right] \!  + \! ... \bigg\} .\label{eq:NLOspectral}
\end{align} 
The on-shell integration over the ultra-relativistic momenta is $\text{d}\Pi_i = \frac{\text{d}^3 p_i}{(2\pi)^3 2 |\mathbf{p}_i|}$, and $P_i=(|\mathbf{p}_i|,\mathbf{p}_i)$. The second line contains Fermi-Dirac distributions. The term shown corresponds to bath-particle scattering. The omitted term contains an off-shell decay process of the vector mediator into a bath-particle pair, which we have found to be sub-dominant compared to the scattering case. By taking the imaginary part of the propagator in the second line to compute the spectral function, and together with the transition elements in Eq.~(\ref{eq:Tgroundstate}), we have for the first time computed the DM bound-state formation cross section via bath-particle scattering ({\bf{bBSF}}) for a massive vector mediator:
\begin{align}
\int \frac{\text{d} \Omega_{\mathbf{k}}}{4\pi} (\sigma_{100}^{\bBSF} \vrel)   = 
 \frac{4  h(\zeta,\xi)\Delta E}{3 \mu^3} 
\times   R_{100}^\bBSF.
\label{eq:FactorOut_bBSF}
\end{align}
The angular average arises in the thermal average of the cross section. We find it here for a computational reason meaningful to already perform. The first term is identical to the one also appearing in the case of mBSF, see Eq.~(\ref{eq:xsectionmbsf}). $R_{100}^\bBSF$ encompasses the details of the bath-particle scattering and is given
in dimensionless coordinates $y\equiv|\mathbf{p}_1|/T$ by:
\begin{align}
&R_{100}^\bBSF \! =
2\times 
\frac{\aD}{2\pi} \(\frac{T}{\Delta E}\)^3
\!\! \int_0^\infty \!\!\! \text{d}y
\frac{1}{e^y + 1} \(1 - \frac{1}{e^{\Delta E/T+y} + 1} \) 
\nn \\
&\times
\left\{
\[ y^2 + \(\frac{\Delta E}{T}+y\)^2 + \frac{\mV^2}{T^2} \]
\ln \[1 + \frac{4y (\Delta E/T+y)}{\mV^2/T^2}\]
\right. \nn \\ &\left.
- \frac{4y (\Delta E/T+y) \[\mV^2/T^2 + (\Delta E/T+2y)^2 \]}
{\mV^2/T^2 + 4 y (\Delta E/T+y)}
\right\} .
\label{eq:R100_bBSF_thermal}
\end{align}
The amount of integrals were reduced analytically down to only one single remaining, without taking further approximations. The factor 2 originates from summing over particle and anti-particle scattering, already contained in Eq.~(\ref{eq:NLOspectral}). We would like to stress the fact that bBSF has no kinematical block, and has an additional temperature and velocity-enhancement factor $(T / \Delta E)^3$ for $T \gg |E_{100}|$.
It was explicitly checked that Eq.~(\ref{eq:FactorOut_bBSF}) is identical to the expression one would obtain in the Boltzmann formalism. The benefit of starting from a proper thermal field theory definition Eq.~(\ref{eq:generalxsection}) is that the contribution of additional interference terms is automatically taken into account. In the Boltzmann framework they are absent and arise from taking instead the imaginary part of the double mediator propagator in Eq.~(\ref{eq:Dyson}). For our model, we find that their contribution for $m_V \gg |E_{100}|$ is negligible. However, notice that the additional interference terms regulate Eq.~(\ref{eq:R100_bBSF_thermal}) in the limit $m_V\rightarrow 0$, by canceling the forward scattering divergence of the bath-particles, as mathematically proven in ref.~\cite{Binder:DivCancel}. In the latter work, it is also shown that the UV-divergent vacuum parts in Eq.~(\ref{eq:NLOspectral}) do not contribute significantly after standard renormalization. This shows the power of Eq.~(\ref{eq:generalxsection}). It contains \emph{all} possible processes up to a certain order in the coupling expansion.

\section{Numerical results and implications}
\label{sec:impl}
\begin{figure}[t!]
\includegraphics[width=1\linewidth]{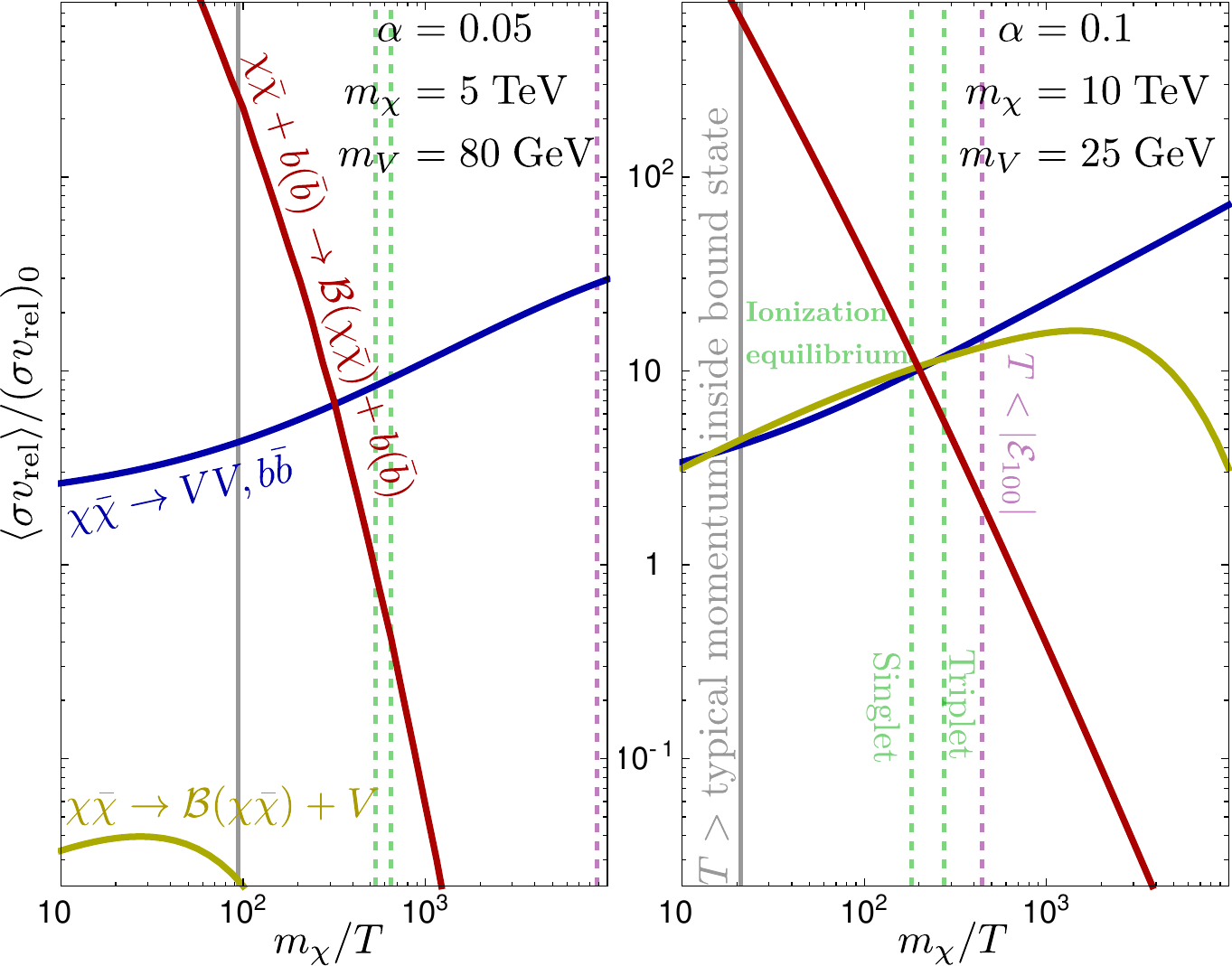} 
\caption[]{\label{fig:CrossSections}
Comparison of thermally averaged cross sections for SE annihilation (blue), ground-state capture via on-shell mediator emission (yellow) and via bath-particle scattering (red) shown. All $\langle\sigma \vrel\rangle$ are normalized to $(\sigma \vrel)_0 \equiv \pi \aD^2/m_\chi^2$.}
\end{figure}
The thermal average of the BSF cross sections via on-shell mediator emission Eq.~(\ref{eq:xsectionmbsf}) and via bath-particle scattering Eq.~(\ref{eq:FactorOut_bBSF}), as well as of the Sommerfeld enhanced annihilation cross section (App.~\ref{app:SE}) are compared in fig.~\ref{fig:CrossSections}. Two different cases are shown where in both $m_V \gtrsim |E_{100}|$. In the left panel, only one bound state close to threshold exists, aiming to parametrically resemble Wino-neutralino DM scenarios. In the right panel, the coupling has similar values as the strong coupling of the SM. The ground state is nearly Coulombic, aiming to provide insight into, e.g., co-annihilation with color-charged particles (neutralino-squark) and multi-TeV WIMPs, typically residing in higher $SU(2)_L$ representations. Overall one can immediately recognize that the results show the proof of concept, i.e., bBSF can entirely dominate over mBSF. In sharp contrast to mBSF, bBSF is kinematically accessible even if $\mV \gtrsim \Delta E$. 

The reason why bBSF is strongly enhanced compared to mBSF, can be qualitatively explained as in the following. First, notice from Eq.~(\ref{eq:FactorOut_bBSF}) that bBSF is enhanced at high temperatures by the large number density of the ultra-relativistic bath particles $n_{b} \propto T^3$. Second, while mBSF is suppressed by the smallness of the dissipated energy $(\sigma^{\mBSF} v_{\text{rel}}) \propto |{\bf p}| = \Delta E $~\cite{Petraki:2016cnz}, bBSF is enhanced by it: $(\sigma^\bBSF v_{\text{rel}})  \propto 1/\Delta E^2$ has a stronger dependence on the inverse relative velocity. One reason for this is that the momentum of the off-shell $V^\mu$ is  not fixed by $\Delta E$, as is in mBSF. These features yield the relative scaling $R^\bBSF \propto 1/\Delta E^3$. An important similarity, however, is that both processes fulfill the same angular momentum selection rules, seen from the factorization in Eq.~(\ref{eq:generalxsection}). Hence, forbidden channels can also not be accessed via higher order processes contained in the mediator spectral function.

The implications of a large bBSF rate on the evolution of the DM density can be seen from the Boltzmann equation, including the lowest bound states as~\cite{vonHarling:2014kha,Ellis:2015vna}:
\begin{align}
\dot{n}_{\chi} + &3Hn_{\chi}  = 
- \left[\langle \sigma^{\text{an}} v_{\text{rel}} \rangle+ W(T) \right]
\left[n_{\chi}^2- (n_{\chi}^{\text{eq}})^2 \right], 
\label{eq:Boltzmann_Approx} 
\\
W \equiv &  
\langle  \sigma^{\BSF}_{100} v_{\text{rel}}  \rangle   \[
\frac{ (1/4) \, \Gamma_{100,\rm S}^{\text{dec}}}
{\Gamma_{100,\rm S}^{\text{dec}} +  \Gamma_{100}^{\rm dis}} 
+
\frac{ (3/4) \, \Gamma_{100,\rm T}^{\text{dec}}}
{\Gamma_{100,\rm T}^{\text{dec}} +  \Gamma_{100}^{\rm dis}} 
\] .
\label{eq:Weff}
\end{align}
The depletion of the DM number density depends on the SE annihilation cross section and the effective cross section $W$. The latter quantity stores all the information of the bound states and consists of the
total thermally averaged BSF cross section, weighted by branching fractions containing the decay rate of spin-singlet and triplet bound states into two $V$ and a pair of bath particles, respectively (see~App.~\ref{app:SE}). The factors 1/4 and 3/4 account for the spin multiplicities. The \emph{dissociation rate} is related via detailed balance to the BSF cross section as $\Gamma_{100}^{\text{dis}} = \langle  \sigma^{\BSF}_{100} v_{\text{rel}}  \rangle (n_{\chi}^{\text{eq}})^2/n_{100}^{\text{eq}}$, where $n_{100}^{\text{eq}}$ is the ground state equilibrium number density with four spin d.o.f.

The effective cross section has two asymptotic regimes:
\begin{equation}
W(T)\simeq \begin{cases*}
       \Gamma_{100}^{\text{dec}}n_{100}^{\text{eq}}/(n_{\chi}^{\text{eq}})^2 & \text{for} $\Gamma_{100,i}^{\text{dec}}\ll \Gamma_{100}^{\text{dis}},$ \\
      \langle\sigma^{\BSF}_{100} v_{\text{rel}}\rangle & \text{for} $\Gamma_{100,i}^{\text{dec}}\gg \Gamma_{100}^{\text{dis}}$,
   \end{cases*}
\end{equation}
where $\Gamma_{100}^{\text{dec}}= (\Gamma_{100,\rm S}^{\text{dec}}+ 3 \Gamma_{100,\rm T}^{\text{dec}})/4 $. A large BSF cross section implies that $\Gamma_{100,i}^{\text{dec}}\ll \Gamma_{100}^{\text{dis}}$ at early times. In this first asymptotic regime, the system is in a phase of Saha \emph{ionization equilibrium} (io.~eq.), where $W$ is i) \emph{independent} of the BSF cross section~\cite{Binder:2018znk}, and ii) \emph{maximal} for a given temperature and bound-state decay rate. The end of the initial io.~eq. phase is marked by the green lines in Fig.~\ref{fig:CrossSections}, where $\Gamma_{100,i}^{\text{dec}} = \Gamma_{100}^{\text{dis}}$ for the spin-singlet and triplet states in our model. Importantly, this transition depends on the value of total BSF cross section contained in $\Gamma_{100}^{\text{dis}}$. In the second asymptotic regime at later times, the number density depletion depends on the actual BSF cross section, since the bound states immediately decay without being dissociated back into the scattering states.

From this discussion one sees that an additional BSF channel can significantly increase the total depletion.  First, a large bBSF contribution implies that io.~eq. is sustained until much lower temperatures than if mBSF was the only capture process. Secondly, bBSF can significantly enhance or entirely dominate the total BSF rate after the end of io.~eq., even if briefly so, as seen in Fig.~\ref{fig:CrossSections}.

The dipole approximation for bBSF breaks down at $T$ larger than the typical momentum inside the bound state, marked by the gray line in Fig.~\ref{fig:CrossSections}. However, since the system is robustly in io.~eq., the actual magnitude of the BSF cross section is irrelevant. It is possible though, that the thermal bath at high T affects the SE annihilation
cross section and the bound state decay rate, which are the only quantities that determine the collision term in Eq.~\eqref{eq:Boltzmann_Approx} in io.~eq.

The impact of the primordial plasma environment on the SE annihilation and bound-state decay rate is already well studied within a different thermal field theory approach~\cite{Kim:2016kxt,Kim:2016zyy,Biondini:2017ufr,Biondini:2018xor,Binder:2018znk,Biondini:2018pwp,Biondini:2018ovz,Kim:2019qix,Biondini:2019int,Biondini:2019zdo}. In particular, Eq.~(\ref{eq:Dyson}) can be resummed in the Hard-Thermal-Loop approximation~\cite{Braaten:1989mz}, giving an effective in-medium potential for a Schr\"odinger-like equation, including real part corrections (e.g., Debye screening), as well as an imaginary thermal width~\cite{Laine:2006ns,Brambilla:2008cx}. A most conservative estimate~\cite{Kim:2016kxt} shows that the latter leads to an entire \emph{melting} of the bound states (see, e.g., fig.~1 in~\cite{Sirunyan:2018nsz} for data) for $T$ to the left of the gray line, where ultra-efficient bath-particle scattering strongly mixes the bound and scattering states.

The formal key-point of this work is that the generalized BSF cross section Eq.~\eqref{eq:generalxsection} \cite{Binder:DivCancel} has overlap with the validity region of the formalism used in~\cite{Kim:2016kxt,Kim:2016zyy,Biondini:2017ufr,Biondini:2018xor,Binder:2018znk,Biondini:2018pwp,Biondini:2018ovz,Kim:2019qix,Biondini:2019int,Biondini:2019zdo}, where the latter is strictly speaking limited to io.~eq.~\cite{Binder:2018znk,Biondini:2019zdo}. Combined, we have achieved a complete procedure to accurately calculate thermal relics in the Early Universe for any temperature regime, ranging from the melting of bound states at high $T$ down to far below their decoupling from ionization equilibrium.

\section{Conclusion}
\label{sec:con}

The Early Universe consists of more than about 100 relativistic d.o.f., which can in principle all contribute to the DM bound-state formation process via particle scattering. In this work, we have analyzed the contribution of only one single species of relativistic fermions, coupled to DM through a mediator at the electroweak mass scale. While in the standard on-shell emission computation, the BSF rate can be highly suppressed for such a case, we have demonstrated that BSF via bath-particle scattering can be highly active. In addition, it was emphasized that the latter can lead to ionization equilibrium, where the effective depletion cross section is maximized for a given temperature. The implications of our results are from the broader literature perspective (based on the on-shell emission), that bound-state effects in WIMP models can become more pronounced during the thermal freeze-out process. Consequently, we conclude that DM could be heavier than previously expected, eventually informing indirect and collider search strategies.

\section*{Acknowledgements}
T.B. was supported by World Premier International Research Center Initiative (WPI), MEXT, Japan.
K.P. was supported by the ANR ACHN 2015 grant (``TheIntricateDark" project), and by the NWO Vidi grant ``Self-interacting asymmetric dark matter". We are thankful to Burkhard Blobel, Peter Cox, Julia Harz and Shigeki Matsumoto for very stimulating discussion, as well as to Satoshi Shirai for checking the manuscript.

\appendix

\section{Bound-state formation cross section}
\label{app:bsf}
This appendix complements ref.~\cite{Binder:DivCancel}, where the generalized BSF cross section Eq.~(\ref{eq:generalxsection}) is derived from pNRQFT in the framework of non-equilibrium quantum field theory. Here, we start from the more commonly known Boltzmann framework and derive Eq.~(\ref{eq:generalxsection}) at the free order in the two-point correlator, where thermal field theory and the Boltzmann approach can be brought together.

In the standard Boltzmann framework, the cross section for describing the conversion of a $\chi \bar{\chi}$ pair into a bound state with quantum numbers $nlm$ via the emission of an on-shell boson $\phi$ (here general), is given by:
\begin{align}
&\sigma_{nlm}^{\BSF} v_{\text{rel}} \! = \! \frac{1}{ 4 g_{\chi} g_{\bar{\chi}} m_{\chi}^2 } \! \int \! \! \frac{\text{d}^3 p}{(2\pi)^3 2 E_{\mathbf{p}}} \frac{\text{d}^3 q}{(2\pi)^3 2 M} \! \nonumber \\ \! & \times \left[1 + f^{\text{eq}}_\phi(E_{\mathbf{p}}) \right]  \slashed{\delta}^4(K_{\chi} + K_{\bar{\chi}} - P - Q)  |\mathcal{M}|^2_{\mathbf{k}, nlm}. \label{eq:boltzmann}
\end{align}
The thermal average for all cross sections in our work follows the standard convention $\langle (...) \rangle= (...)$ if $(...)$ is velocity independent. Final state integration is over the emitted boson four momenta $P=(P^0,\mathbf{p})$, and over the dark matter bound state four momentum $Q=(Q^0,\mathbf{q})$. The four momenta of the initial DM two-body state, as well as of the final bound state with mass $M$ are described non-relativistically. The slashed convention is $\slashed \delta^n =(2 \pi)^n  \delta^n$. The squared amplitude $|\mathcal{M}|^2_{\mathbf{k},nlm} $ for the bound-state formation process in Born approximation is \emph{summed} over initial and final state internal degrees of freedom.

To allow for general dispersion relations of the mediator at the end, the on-shell momentum integration is first rewritten into four-momentum integration as
\begin{align}
\!\!\!\!\!\!&\int \frac{\text{d}^3 p}{(2\pi)^3 2 E_{\mathbf{p}}}\left[1 + f^{\text{eq}}_\phi(E_{\mathbf{p}}) \right] =  \nonumber \\ \!\!\!\!\!\!&\int\!\! \frac{\text{d}^4 P}{(2\pi)^4} \! \! \left[1\!+\!f^{\text{eq}}_\phi(P^0)\right]\!\!\left[ \theta(P^0) \text{sgn}(P^0) \slashed\delta(P^2-m_\phi^2) \right], \\
\!\!\!\!\!\!& \int \frac{\text{d}^3 q}{(2\pi)^3 2 M} \nonumber = \\ \!\!\!\!\!\!&\int \frac{\text{d}^4 Q}{(2\pi)^4} \left[ \frac{1 }{2 M} \slashed\delta(Q^0-M- |\mathbf{Q}|^2/(2M)) \right].
\end{align}
One can recognize that the mediator and bound-state dispersion relation are here on the mass shell.
Within this notation the BSF cross section can be written as:
\begin{align}
&\sigma_{nlm}^{\BSF} v_{\text{rel}} = \frac{1}{ N^2} \nonumber \\
& \times \! \int \! \frac{\text{d}^4 P}{(2\pi)^4} \left[1+f^{\text{eq}}_\phi(P^0)\right]\!\left[ \theta(P^0) \text{sgn}(P^0) \slashed\delta(P^2-m_\phi^2) \right] \nonumber \\
&\times \slashed\delta\left(P^0-\Delta E - \frac{\mathbf{K^2}}{4m_{\chi}} + \frac{(\mathbf{K}-\mathbf{p})^2}{4m_{\chi}}\right) |\mathcal{M}|^2_{\mathbf{k}, nlm}. \label{eq:tempxsection}
\end{align}
Here, we defined $N=(g_{\chi} g_{\bar{\chi}}4m_{\chi}^2 2M)^{1/2}$, see Eq.~(\ref{eq:tdef}), and used the four-momentum conserving delta function in Eq.~(\ref{eq:boltzmann}) to perform the $Q$ integration. Furthermore, we switched to the DM CM-momenta coordinates where $\mathbf{K}\equiv \mathbf{k}_\chi + \mathbf{k}_{\bar{\chi}}$ and $\mathbf{k} \equiv (\mathbf{k}_\chi -  \mathbf{k}_{\bar{\chi}})/2$ with $|\mathbf{k}|= \mu v_{\text{rel}}$. In the \emph{zero momemtum-recoil approximation}, the delta function can be estimated as: \footnote{In the zero-momentum-recoil approximation it is assumed that $\mathbf{K} \gg \mathbf{p}$, which is justified since typical DM CM-momentum $\mathbf{K}$ ($\sim \sqrt{m_{\chi}T}$) is much larger than the momentum of the emitted particle $\mathbf{p}$  ($\sim \Delta E \sim \text{Max}[T,|E_{100}|]$) in the bound-state formation process. Already around the freeze-out temperature, where DM enters the non-relativistic regime this holds, and for temperature larger than $ \sim \alpha^2 |E_{100}|$. This captures the entire region of the freeze-out we are interested in, since DM decouples from the bound states when temperature is already less than the ground state energy, i.e., $T\sim |E_{100}| \gg \alpha^2 |E_{100}|$. For electroweak couplings we have checked the zero-recoil approximation in the high temperature regime $m_{\chi}/T=10$ and noticed that the difference is about only one per-cent compared to the exact result (but still assuming dipole approximation). Also note that ref.~\cite{Binder:DivCancel} assumes the zero-momentum-recoil approximation to be valid from the beginning, since the derivation is based on pNRQFT.
}
\begin{align}
\! \! \! \delta\left( \! P^0-\Delta E - \frac{\mathbf{K^2}}{4 m_{\chi}} + \frac{(\mathbf{K}-\mathbf{p})^2}{4m_{\chi}}\right) \! \sim \delta\left(P^0-\Delta E\right)
\end{align}
and the BSF cross section can be brought into the form:
\begin{align}
\!\!\!&\sigma_{nlm}^{\BSF} v_{\text{rel}} =\frac{1}{ N^2} \int \frac{\text{d}^4 P}{(2\pi)^4} \slashed\delta\left(P^0-\Delta E \right) \nonumber \\& \times \! \left[1+f^{\text{eq}}_\phi(P^0)\right] \!  \left[\text{sgn}(P^0) \slashed\delta(P^2-m_\phi^2) \right] \! |\mathcal{M}|^2_{\mathbf{k}, nlm}.\label{eq:intermediatecr}
\end{align}
Here, the theta function $\theta(P^0)$ was dropped, since the delta function in the first line already ensures positivity of the emitted energy. Assuming a massive vector field, one can decompose the amplitude as
\begin{align}
\!\!\!|\mathcal{M}|^2_{\mathbf{k},nlm} \! &= \! \sum_{\sigma} \epsilon_{\mu}(P,\sigma)\epsilon_{\nu}^{\star}(P,\sigma) \sum_{\text{spins}} \mathcal{M}^{\mu}_{\mathbf{k},nlm} \mathcal{M}^{\nu \star}_{\mathbf{k},nlm} \nonumber \\
&= \! - \! \left(g_{\mu \nu}-\frac{P^{\mu}P^{\nu}}{m_{\phi}^2}\right) \! \sum_{\text{spins}}  \mathcal{M}^{\mu}_{\mathbf{k}, nlm} \mathcal{M}^{\nu \star}_{\mathbf{k},nlm},
\end{align}
where $\epsilon$ is the polarization vector. Note that in our model, $\mathcal{M}^{\mu}_{\mathbf{k}, nlm}$ here is identical to what has been defined already in Eq.~(\ref{eq:vertex}). By factoring out the polarization tensor dependence from the squared matrix element in Eq.~(\ref{eq:intermediatecr}), one may notice the lesser two-point correlation function in thermal equilibrium Eq.~(\ref{eq:KMS}) in the free limit:
\begin{align}
&D^{-+,0}_{\mu \nu}(P^0,\mathbf{p}) =\left[1+f^{\text{eq}}_\phi(P^0)\right] \times \nonumber  \\ & \left[\left(-g_{\mu \nu}+\frac{P^{\mu}P^{\nu}}{m_{\phi}^2}\right)\text{sgn}(P^0) \slashed\delta(P^2-m_\phi^2) \right].
\end{align}
After integrating Eq.~(\ref{eq:intermediatecr}) over $P^0$, and replacing the free lesser function by interacting one, Eq.~(\ref{eq:generalxsection}) is recovered. The justification of this replacement from first principles can be found in ref.~\cite{Binder:DivCancel}. Here, we would like to argue from the Kadanoff-Baym equation that this ``upgrade" is expected. As derived in ref.~\cite{Beneke:2014gla}, the DM collision term in the forward direction can be determined through the DM self-energy component $\Sigma^{-+}$. In the 1-PI formalism, the $\Sigma^{-+}$ contains interacting $D^{-+}_{\mu \nu}$. This is another indication that the upgrade from vacuum correlations in the Boltzmann formalism to the interacting correlation functions in the thermal field theory approach is indeed justified. Finally we would like to remark that this scheme of derivation presented in this Appendix also holds for scalar mediators, where at the end, one has to just drop the Greek indices in Eq.~(\ref{eq:generalxsection}).

\section{Retarded self-energy expression}
\label{app:selgenergy}
The retarded self-energy is defined in terms of greater and lesser self-energies as
\begin{align}
\!\!\!\!\!\!\Pi_{\mu \nu}^R(x-y) \! &= \! \theta(x^0-y^0) \!\! \left[ \Pi^{-+}_{\mu \nu}(x-y) \! - \!  \Pi^{+-}_{\mu \nu}(x-y) \right],\label{eq:retardeddef} \\
\!\!\!\!\!\!\Pi^{+-}_{\mu \nu}(x-y)&= g^2 \Tr[ \gamma_{\mu} B^{+-}(x-y) \gamma_{\nu} B^{-+}(y-x)],\\
\!\!\!\!\!\!\Pi^{-+}_{\mu \nu}(x-y)&= g^2 \Tr[ \gamma_{\mu} B^{-+}(x-y) \gamma_{\nu} B^{+-}(y-x)],
\end{align}
where the two-point functions of the fermionic bath-particles are
\begin{align}
B^{-+}_{ij}(x-y)&\equiv\langle b_i(x) \bar{b}_j(y) \rangle, \\ B^{+-}_{ij}(x-y)&\equiv-\langle\bar{b}_j(y)  b_i(x) \rangle,
\end{align}
and their free Fourier transforms are in equilibrium
\begin{align}
\!\!\!\!\!\!B^{+-}(K)&= -\slashed{K}(2\pi)\delta(K^2)\left[-\theta(-K^0) + f_b^{\text{eq}}(|K^0|) \right],\\
\!\!\!\!\!\!B^{-+}(K)&= -\slashed{K}(2\pi)\delta(K^2)\left[-\theta(+K^0) + f_b^{\text{eq}}(|K^0|) \right].
\end{align}
The Fourier transform of the retarded self-energy is then
\begin{align}
\! \! \!  \! \! \! \! \! \! \! \! \! &\Pi^{R}_{\mu \nu}(P)=  g^2 \int \text{d}\Pi_1 \text{d}\Pi_2  \Tr[\gamma_{\mu} \slashed{P}_1 \gamma_{\nu} \slashed{P}_2] \times  \\
\! \! \! \bigg\{ \! &  \frac{i  \slashed{\delta}^3(\mathbf{p}+\mathbf{p}_1-\mathbf{p}_2)}{P^0 +|\mathbf{p}_1|-|\mathbf{p}_2| + i \epsilon} \! \left[ f^{\text{eq}}_b(|\mathbf{p}_1|) \! - \! f^{\text{eq}}_{b}(|\mathbf{p}_2|) \right] \!  + \nonumber\\
\! &  \frac{i  \slashed{\delta}^3(\mathbf{p}+\mathbf{p}_2-\mathbf{p}_1)}{P^0 +|\mathbf{p}_2|-|\mathbf{p}_1| + i \epsilon} \! \left[ f^{\text{eq}}_b(|\mathbf{p}_2|) \! - \! f^{\text{eq}}_{b}(|\mathbf{p}_1|) \right] \!  + \nonumber\\ 
\! &  \frac{i  \slashed{\delta}^3(\mathbf{p}-\mathbf{p}_1-\mathbf{p}_2)}{P^0-|\mathbf{p}_1|-|\mathbf{p}_2| + i \epsilon} \! \left[1- f^{\text{eq}}_b(|\mathbf{p}_1|) \! - \! f^{\text{eq}}_{b}(|\mathbf{p}_2|) \right] \!  + \nonumber\\
\! &  \frac{i  \slashed{\delta}^3(\mathbf{p}+\mathbf{p}_1+\mathbf{p}_2)}{P^0 +|\mathbf{p}_1|+|\mathbf{p}_2| + i \epsilon} \! \left[-1+ f^{\text{eq}}_b(|\mathbf{p}_1|) \! + \! f^{\text{eq}}_{b}(|\mathbf{p}_2|) \right] \!  \bigg\} .\nonumber
\end{align} 
The first and second term inside the brackets corresponds to particle and anti-particle scattering, when taking the imaginary part of the propagator. Third and fourth line corresponds to off-shell decay into a bath-pair, as well as the reverse process. The latter does not contribute, since in our case $P^0=\Delta E$ is positive, leading to the fact that the cut of the propagator is zero. The following identity was used for computing Eq.~(\ref{eq:R100_bBSF_thermal}):
\begin{align}
&\left[1+f_V^{\text{eq}}(\Delta E)\right]\left[f_b^{\text{eq}}(|\mathbf{p}_1|)-f_b^{\text{eq}}(|\mathbf{p}_1|+\Delta E)\right]= \nonumber \\ & f_b^{\text{eq}}(|\mathbf{p}_1|) \left[1-f_b^{\text{eq}}(\Delta E+|\mathbf{p}_1|)\right],
\end{align}
which makes the particle scattering more apparent. The Bose-enhancement factor comes from Eq.~(\ref{eq:KMS}). Similar identities can also be derived for the off-shell decay.

\section{States convention and overlap integral}
\label{app:states}
In terms of particle and anti-particle creation operators, the states in Eq.~(\ref{eq:vertex}) follow the convention:
\begin{align}
\!\!\! \!\! &\ket{\mathcal{B}_{nlm}} \! = \! \sqrt{2M} \! \!\int \!\! \frac{\text{d}^3p^{\prime}}{(2 \pi)^3} \psi_{nlm}(\mathbf{p}^{\prime}) a^{s^{\prime}\dagger}_{Q/2+p^{\prime}} b^{r^{\prime}\dagger}_{Q/2-p^{\prime}} \ket{0}, \\
\!\!\! \!\!&\ket{\mathcal{S}_{\mathbf{k}}} \!= \!\sqrt{4 m_{\chi}^2 } \!\int \!\frac{\text{d}^3p^{\prime \prime}}{(2 \pi)^3} \psi_{\mathbf{k}}(\mathbf{p}^{\prime \prime}) a^{s\dagger}_{K/2+p^{\prime \prime}} b^{r \dagger}_{K/2-p^{\prime \prime}} \ket{0} .
\end{align}
$Q$ is the momentum of the bound and $K$ is the CM-momentum of the scattering state.
Due to the spin-sum in Eq.~(\ref{eq:generalxsection}), we consider for simplicity an unpolarized BSF cross section which is however not necessary to do. The mass norms from the states cancel out in Eq.~(\ref{eq:tdef}).
The reduced spin-independent two-body wave functions encode the plane-wave distortion due to the long-range interactions. They have energy Eigenvalues ${\cal E}_{\bf k} = {\bf k^2}/(2\mu) = \mu \vrel^2/2$ and negative binding energy ${\cal E}_{nlm}$, and are normalized to plane waves in the case of scattering states, while for the bound states the spatial integral over the absolute squared wave function is normalized to unity. The scattering state is a coherent sum over all partial waves.

In Eq.~(\ref{eq:Tgroundstate}) we defined $h(\zeta,\xi)$ in terms of the overlap integral:
\begin{align}
\frac{\sqrt{h(\zeta,\xi)}}{\kappa^{1/2}\mu} \frac{\hat{k}^i }{\Delta E} \equiv \abs{ \int \text{d}^3 r \; \psi^{\star}_{100}(\mathbf{r}) r^i \psi_{\mathbf{k}}(\mathbf{r}) },
\end{align}
where $\kappa \equiv \mu \alpha$ is the Bohr-momentum. 
We have extracted a specific combination of dimensional quantities to make $h (\zeta, \xi)$ dimensionless.
For convention, the analytic expression for $h(\zeta,\xi)$ in the Coulomb limit $\xi\rightarrow \infty$ is given by~\cite{Petraki:2016cnz}:
\begin{align}
h(\zeta, \infty) &= 2^6 \pi S(\zeta) \frac{\zeta^6}{(1+ \zeta^2)^3} e^{-4 \zeta \text{acot}\zeta}, \\
S(\zeta)&=\left( \frac{2\pi \zeta}{1-e^{-2 \pi \zeta}} \right).
\end{align}

\section{Sommerfeld enhanced annihilation and decay rate}
\label{app:SE}

The $\chi \bar{\chi}$ pairs can annihilate into $VV$ or $b\bar{b}$. The spin-averaged $s$-wave velocity-weighted cross sections are
\begin{align}
\sigma^{\text{ann}}_{\chi \bar{\chi} \to VV}  v_{\text{rel}} = 
\sigma^{\text{ann}}_{\chi \bar{\chi} \to b\bar{b}}  v_{\text{rel}}
&= \frac{\pi\alpha^2}{4\mu^2} \, |\psi_{\mathbf{k},l=0}(r=0)|^2 ,
\label{eq:Annihilation}
\end{align}
where the first factor is the tree-level value, and the squared wavefunction is the Sommerfeld enhancement. The total annihilation cross section appearing in Eq.~\eqref{eq:Boltzmann_Approx} is 
$\sigma^{\text{ann}}  v_{\text{rel}} = \sigma^{\text{ann}}_{\chi \bar{\chi} \to VV}  v_{\text{rel}} + \sigma^{\text{ann}}_{\chi \bar{\chi} \to b\bar{b}}  v_{\text{rel}}$. Note that to working order, only the spin singlet state contributes to $\chi \bar{\chi} \to VV$, and only the spin triplet state contributes to  $\chi \bar{\chi} \to b\bar{b}$.

Similarly to annihilation of unbound pairs, the decay rates of the $\ell=0$ bound states factorize into the corresponding tree-level $s$-wave velocity-weighted annihilation cross sections, and the bound-state squared wavefunctions evaluated at the origin. Here, the tree-level cross sections must be averaged with respect to the bound-state degrees of freedom only. Thus, 
\begin{align}
\Gamma_{n00,S}^{\text{dec}} &= 
\frac{\pi\alpha^2}{\mu^2} \, |\psi_{n00}(r=0)|^2, 
\\
\Gamma_{n00,T}^{\text{dec}} &= 
\frac{\pi\alpha^2}{3\mu^2} \, |\psi_{n00}(r=0)|^2.
\end{align}
Note that $|\psi_{n00}|^2$ carries mass dimension three.

\bibliography{Bibliography.bib}

\end{document}